\documentclass[aps,pra,twocolumn,superscriptaddress]{revtex4-2}

\usepackage{amsmath,amssymb}
\usepackage{siunitx}
\usepackage{braket}
\usepackage{graphicx}
\usepackage{xcolor}

\def\U#1{{\rm #1}} 
\def\u#1{_{\rm #1}}
\newcommand{\od}[2]{\frac{\mathrm{d} #1}{\mathrm{d} #2}}
\newcommand{\ketbra}[2]{| #1 \rangle \langle #2 |}

\def\tr{{\rm tr}}

\begin{document}

\title{A quantum frequency conversion hub interfacing with DWDM networks}

\author{Masatake Yamada}
\affiliation{Graduate School of Engineering Science, The University of Osaka, Osaka 560-8531, Japan}

\author{Kurama Hirano}
\affiliation{Graduate School of Engineering Science, The University of Osaka, Osaka 560-8531, Japan}

\author{Masahiro Yabuno}
\affiliation{Advanced ICT Research Institute, National Institute of Information and Communications Technology (NICT), Kobe, 651-2492, Japan}

\author{Shigehito Miki}
\affiliation{Advanced ICT Research Institute, National Institute of Information and Communications Technology (NICT), Kobe, 651-2492, Japan}

\author{Tsuyoshi Kodama}
\affiliation{Hamamatsu Photonics K.K., Japan}

\author{Hideki Shimoi}
\affiliation{Hamamatsu Photonics K.K., Japan}

\author{Takashi Yamamoto}
\affiliation{Graduate School of Engineering Science, The University of Osaka, Osaka 560-8531, Japan}
\affiliation{\mbox{Center for Quantum Information and Quantum Biology, The University of Osaka, Osaka 560-0043, Japan}}

\author{Rikizo Ikuta}
\email{ikuta.rikizo.es@osaka-u.ac.jp}
\affiliation{Graduate School of Engineering Science, The University of Osaka, Osaka 560-8531, Japan}
\affiliation{\mbox{Center for Quantum Information and Quantum Biology, The University of Osaka, Osaka 560-0043, Japan}}

\begin{abstract}
Interconnecting heterogeneous quantum systems is an important step toward realizing the quantum internet. 
We propose a quantum network hub that interfaces local quantum devices with dense wavelength-division multiplexing (DWDM) networks in the telecom band via quantum frequency conversion (QFC) with frequency-channel selectivity. 
We show that standard periodically poled lithium niobate waveguides used for QFC exhibit a dispersion sweet spot, for example around the 780\,nm band, enabling wide tunability of the pump wavelength while maintaining phase matching. 
Experimentally, we demonstrate the network hub by implementing a channel-selective and polarization-insensitive QFC from 780\,nm to telecom wavelengths around 1540\,nm. 
We achieve a pump tuning range of 2\,THz and successfully distribute polarization-encoded single photons into 16 frequency channels on the ITU-T DWDM grid with 25\,GHz channel spacing, while preserving the quantum information. 
These results position the QFC-based hub as a versatile backbone for connecting a wide range of quantum devices, spanning both photonic and matter-based systems, across frequency-multiplexed telecom networks.
\end{abstract}

\maketitle

\section{Introduction}
Quantum communication among distantly located quantum devices has become an active and rapidly growing area of research in recent years, enabling applications
such as distributed quantum computation,
quantum repeaters, 
quantum key distribution,
networked quantum sensing,
and precise clock synchronization~\cite{Wehner2018,Azuma2023}. 
In demonstrations for long-distance quantum communication over optical fibers~\cite{Yu2020,vanLeent2022,Krutyanskiy2024,Knaut2024,Liu2024,Kucera2024,Stolk2024,Liu2026,Lu2026},
quantum frequency conversion~(QFC) has emerged as a key technique that converts photons emitted from quantum devices at unique wavelengths into telecom-wavelength photons suitable for fiber-optic communication, while preserving quantum information~\cite{Ikuta2011}. 
In such experiments, not all platforms convert photons to the same telecom wavelength,
owing to practical constraints in QFC implementation such as signal-to-noise ratio and preparation of pump light for QFC. 
However, future quantum networks aimed at the quantum internet~\cite{Kimble2008,Wehner2018,Awschalom2021,Azuma2023} will involve multiple users and nodes employing heterogeneous matter-based quantum systems~\cite{Maring2017,Craddock2019,Wang2026}, such as neutral atoms, trapped ions, solid-state spins, semiconductor quantum dots, and superconducting circuits. To establish entanglement among such diverse systems via Bell-state measurements between indistinguishable photons, reconfigurable photonic switching and routing technologies integrating QFC with wavelength-division multiplexing~(WDM) will be important. 

Related to the functionalities,
there have been several experimental works~\cite{Wang2021,Fisher2021,Tang2024,Murakami2025,Von2025,Arizono2026,Murakami2026} and proposals~\cite{Lee2022,Chen2023,Miller2025,Shapourian2025,Zhao2025,Lukens2025} for quantum network architecture with QFCs. 
In particular, Ref.~\cite{Arizono2026} demonstrated widely tunable QFC capable of frequency switching across more than 100 channels, assuming a dense WDM grid with a \SI{25}{GHz} channel spacing.
In the work, channel-selective QFC~(CS-QFC) was successfully demonstrated
using a commonly employed periodically poled lithium niobate~(PPLN) as a second-order nonlinear optical medium,
enabling QFC from a fixed wavelength of \SI{780}{nm} to telecom wavelengths around \SI{1540}{nm} by appropriately switching the pump wavelength for each conversion process.
However, the fundamental physical origin of the wide tunability and its potential applicability
to a broader range of quantum systems emitting photons at different wavelengths remained unexplored. 

\begin{figure}[t]
 \begin{center}
      \scalebox{0.45}{\includegraphics{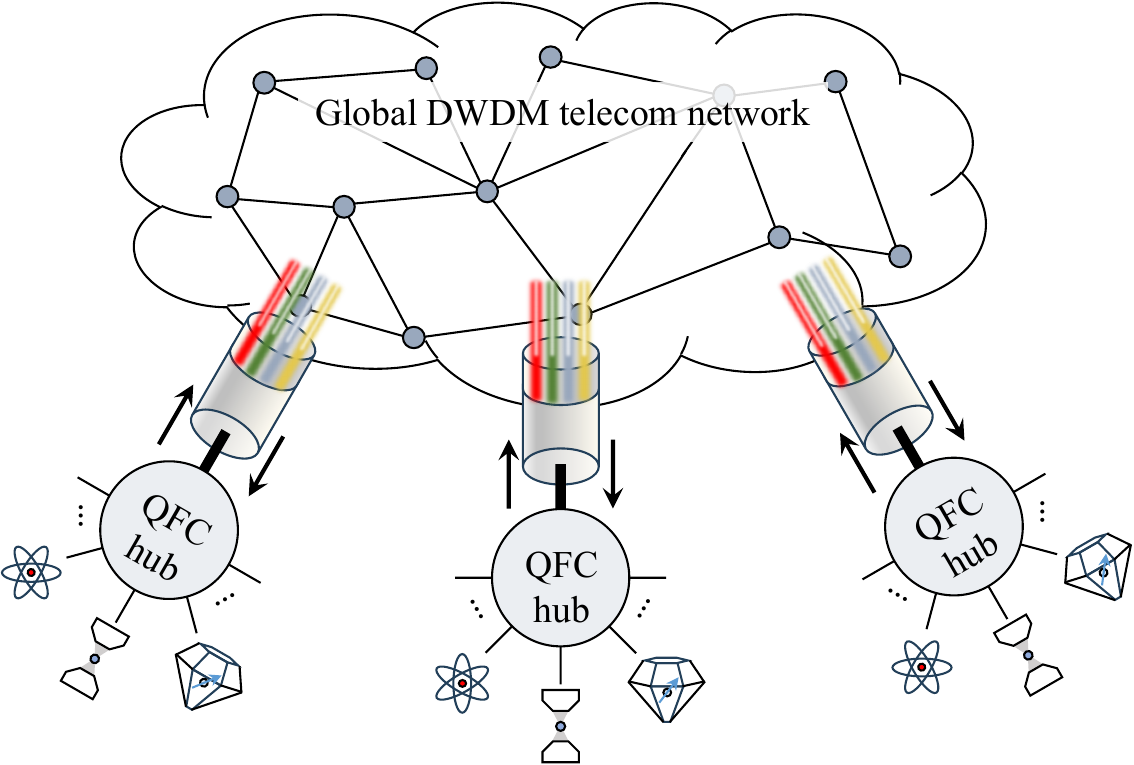}}   
      \caption{
        Concept of QFC hub as an interface 
        between local quantum devices and DWDM telecom network.
    }
    \label{fig:concept}
 \end{center}
\end{figure}
In this work, we reveal that for CS-QFCs interfacing with telecom wavelengths,
wavelengths near the telecom second-harmonic region, including around \SI{780}{nm}, 
satisfy a specific group-velocity condition in standard PPLN waveguides, yielding 
a broad phase-matching bandwidth that is extremely insensitive to variations in the pump and corresponding converted wavelengths. 
This enables broadband tunability of CS-QFC that is difficult to achieve with other common quantum devices interacting with photons at shorter wavelengths such as \SI{422}{nm} for $\U{Sr}^{+}$~\cite{Wright2018,Main2025}
and \SI{393}{nm} for $\U{Ca}^{+}$~\cite{Yang2026}. 
Based on the analytical results, we propose a QFC-based network hub architecture 
interfacing a favorable wavelength band, represented by \SI{780}{nm}, with DWDM networks, 
as shown in Fig.~\ref{fig:concept}. 
Rather than developing separate and narrowly tunable interfaces for individual quantum species, 
this architecture converts their diverse emitted photons into a favorable wavelength band, 
fully leveraging the multiplexing capability of WDM networks.
Combined with a quantum processor equipped with an optical interface,
this QFC-based approach positions the interface not merely as a standard photonic interface,
but as a potential functional quantum repeater node. 
For example, \SI{780}{nm} is compatible with the $\U{D}_2$ line of rubidium~(Rb) atoms,
making it well suited for integration with optically interfaced Rb-based quantum processors~\cite{Young2022,Covey2023,Sinclair2025,Baranes2025,Maeda2025,Safari2026}. 

To realize a practical quantum network hub, the interface must preserve the polarization qubit, 
a critical feature not demonstrated in previous CS-QFC implementations. 
To address this, we present a polarization-insensitive QFC~(PIQFC) with wide tunability. 
In the experiment, we demonstrate a quantum network hub based on CS-PIQFC, 
enabling conversion from \SI{780}{nm} to telecom wavelengths. 
Routing the converted photons through a 16-channel demultiplexer~(DeMux) with a channel spacing of \SI{25}{GHz} on the ITU-T DWDM grid, we achieve a process fidelity of over \SI{75}{\%} in every channel, clearly surpassing the classical limit.

\section{780 nm node as a flexible mediator to WDM network}
The CS-PIQFC in this paper is based on the DFG process in the second-order nonlinear optical interaction. 
It operates as a $1\times N$ QFC switch~\cite{Arizono2026}, 
which means that while an angular frequency $\omega\u{s}$ of the input photon is fixed,
multiple output frequency channels are available. 
In each round of QFC, the pump light at a single frequency $\omega\u{p}$ is used,
and the frequency of the signal photon is converted into a specific target frequency $\omega\u{c}(=\omega\u{s}-\omega\u{p})$.
When the conversion efficiencies for the horizontal~(H) and vertical~(V) polarization are identical, 
the polarization of the signal photon is preserved through the QFC process~\cite{Ikuta2018,Bock2018}. 
By tuning the pump frequency $\omega\u{p}$ in each round, the converted frequency is correspondingly determined, which effectively equips the PIQFC with wavelength-selective switching capability.

In the above CS-PIQFC, the tuning range of the pump and the corresponding converted frequencies is determined by the phase-matching condition. 
In conventional QFCs using PPLN waveguides, the first-order quasi-phase-matching~(QPM) is employed, and therefore the phase mismatch is given by 
\begin{align}
\Delta k = k(\omega\u{s}) - k(\omega\u{p}) - k(\omega\u{c}) - \frac{2\pi}{\Lambda}, 
\end{align}
where $k(\omega)$ is the wave number of photons at angular frequency $\omega$ 
and $\Lambda$ is the poling period of the device. 
The condition $\Delta k = 0$ corresponds to the perfect phase matching among the signal, pump, and converted light. 

We first consider the tunability of the converted wavelengths in the process of QFC from $\lambda\u{s0}=$\SI{780}{nm} to a telecom wavelength at  $\lambda\u{c0}=$\SI{1540}{nm} using a pump light around $\lambda\u{p0}\sim$\SI{1580}{nm} determined by the energy conservation $\omega\u{p0}=\omega\u{s0}-\omega\u{c0}$, 
together with $\omega = 2\pi c/\lambda$, where $c$ is the speed of light. 
To calculate the refractive indices for the wave vectors, we use the Sellmeier equation and parameters given in Ref.~\cite{Jundt1997}, 
assuming type-0 QPM, where all interacting lights are extraordinary polarized. 
With the temperature fixed at $T=$48\,${}^\circ$C, 
we determine the poling period $\Lambda$ such that 
$\Delta k(\lambda\u{s0}, \lambda\u{c0}, \lambda\u{p0}) =0$ is satisfied. 
Using the fixed values of $\lambda\u{s0}$, $T$, and $\Lambda$, 
we evaluate the phase mismatch $\Delta k$ 
as a function of $\lambda\u{c}$ and corresponding $\lambda\u{p}$. 

\begin{figure}[t]
      \scalebox{1.05}{\includegraphics{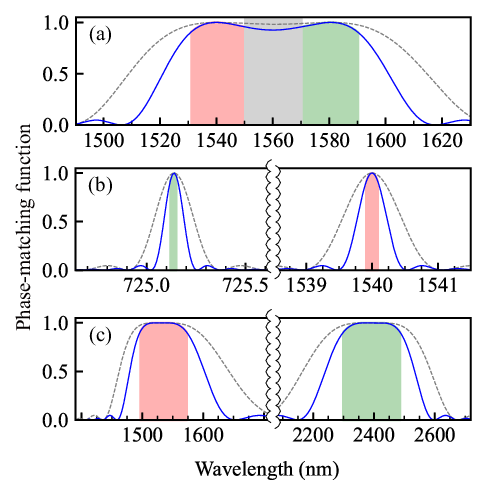}}
      \caption{
        Phase-matching functions for QFCs to $\lambda\u{c0}=\SI{1540}{nm}$ using PPLN waveguides.
        The blue and grey curves correspond to the crystal lengths of $L=\SI{40}{mm}$ and $\SI{20}{mm}$, respectively. 
        The pink and green regions indicate the converted and corresponding pump wavelengths, respectively.
        $\lambda\u{s0}=$ (a) \SI{780}{nm}, (b) $\SI{493}{nm}$, and (c) \SI{934}{nm}. 
        Note the wavelength scales on the the horizontal axes are significantly different. 
      }
    \label{fig:780}
\end{figure}
From the estimated value of $\Delta k$, we obtain the phase-matching function~\cite{Boyd2008},
which determines the maximum conversion efficiency of QFC, as 
\begin{align}
|h(\Delta k)|^2 = \U{sinc}^2 \left(\Delta k \frac{L}{2} \right),  
\end{align}
where $L$ is the length of the PPLN waveguide.
In the phase-matching function shown in Fig.~\ref{fig:780}~(a), we find two peaks, 
which are located at 
$(\lambda\u{c},\lambda\u{p})=(\lambda\u{c0},\lambda\u{p0})$ and $(\lambda\u{p0},\lambda\u{c0})$. 
The tuning range depends on the allowable degradation of the conversion efficiency. 
In this analysis, 
we define the tuning range as the range over which the efficiency remains above \SI{90}{\%} of the maximum value. 
We additionally consider a practical constraint. 
For a high signal-to-noise-ratio QFC in the wavelength regime, 
the condition $\lambda\u{c} < \lambda\u{p}$ is required to avoid contamination by Stokes Raman scattering induced by the strong pump light~\cite{Ikuta2011}. 
Furthermore, even if this condition is satisfied, 
when the pump and converted wavelengths become too close, 
anti-Stokes Raman scattering introduces noise photons, making QFC difficult. 
According to Ref.~\cite{Arizono2026}, high-quality QFC is feasible at least up to a converted wavelength of about $\lambda\u{c}=$\SI{1550}{nm}~(corresponding to pump wavelength~$\lambda\u{p}\sim$\SI{1570}{nm}), which we take as the practical cutoff. 
Under the constraints, from Fig.~\ref{fig:780}~(a),
the effective tuning range of QFC from $\lambda\u{s0}=$\SI{780}{nm} to the telecom wavelengths
centered at $\lambda\u{c0}=$\SI{1540}{nm}
is estimated to be \SI{19.5}{nm} for $L=\SI{40}{mm}$, 
which is in good agreement with the reported value in Ref.~\cite{Arizono2026}
and \SI{32.2}{nm} for $L=\SI{20}{mm}$. 
This means that the number of frequency multiplexing is about 100 and 160 
for $L=\SI{40}{mm}$ and \SI{20}{mm}, respectively, with the channel spacing of \SI{25}{GHz}. 

To highlight the uniqueness of the \SI{780}{nm} band, 
we performed a similar simulation of the tunability of QFCs
from various values of $\lambda\u{s0}$ to $\lambda\u{c0}=\SI{1540}{nm}$. 
An example of the phase-matching function for $\lambda\u{s0}=\SI{493}{nm}$, 
which corresponds to the wavelength of a photon emitted from Ba$^+$,  
is shown in Fig.~\ref{fig:780}~(b). 
In contrast to the broad phase-matching curve shown in Fig.~\ref{fig:780}~(a), 
we see that the conversion of \SI{493}{nm} photons is confined to a narrow peak with a bandwidth of \SI{0.2}{nm}. 
The striking contrast provides the basis for the \SI{780}{nm} quantum network hub architecture. 
By first connecting the \SI{493}{nm} node to the \SI{780}{nm} interface~\cite{Hannegan2021,Saha2023}, 
it becomes possible to access a wide range of DWDM channels. 

\begin{figure}[t]
 \begin{center}
    \scalebox{1}{\includegraphics{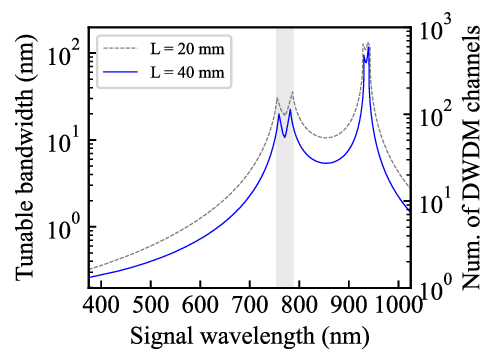}}    
    \caption{
    Tunable bandwidth vs wavelength of the input photon 
    when the center of the target wavelength after QFC is \SI{1540}{nm}. 
    To avoid Raman noise, we evaluate the bandwidth under the condition that the pump and converted wavelengths are separated by \SI{20}{nm}~(gray region), independent of the signal wavelength.
    Channel spacing of the DWDM is assumed to be \SI{25}{GHz}. 
    }
    \label{fig:Cband}
 \end{center}
\end{figure}
Following a similar procedure, we calculated the tuning range of CS-QFC for arbitrary signal wavelengths,
with the center of the target telecom wavelength fixed at $\lambda\u{c0}=\SI{1540}{nm}$.
From the result shown in Fig.~\ref{fig:Cband}, 
the calculated tunability exhibits two prominent maxima around \SI{780}{nm} and \SI{934}{nm}, the origin of which will be discussed later. 
From the viewpoint of tunable range alone, the \SI{934}{nm} band appears advantageous. 
However, practical implementation requires consideration of the availability of the pump laser sources. 
For DFG into the telecom C-band, converting a \SI{934}{nm} photon requires a pump wavelength of approximately $\SI{2350}{nm}$. 
High-performance lasers and amplifiers in this mid-infrared range are currently less accessible and more complex to operate than standard telecom components. 
In contrast, the QFC of \SI{780}{nm} photons utilizes a pump wavelength around $\SI{1580}{nm}$~(telecom L-band), which benefits from the well-established telecom technologies. 
Consequently, we identify the \SI{780}{nm} band as the most practical and deployable candidate for a quantum network hub for connection to DWDM networks.

We explain the physical origin of the two peaks of the tunability shown in Fig.~\ref{fig:Cband}. 
For the peak around \SI{780}{nm}, the wide tunability can be understood from the fact that 
\SI{780}{nm} is the second harmonic of \SI{1560}{nm},
which lies near the midpoint between the target wavelength \SI{1540}{nm} and its corresponding pump wavelength \SI{1580}{nm}. 
To see this, we consider that the perfect QPM condition $\Delta k=0$ is satisfied for $\omega\u{s0}$, $\omega\u{c0}$ and $\omega\u{p0}(=\omega\u{s0}-\omega\u{c0})$.
Under the condition, 
when the frequencies relevant to the QFC are changed as 
$\omega\u{c0}\rightarrow \omega\u{c0}+\Delta\omega$ and 
$\omega\u{p0}\rightarrow \omega\u{p0}-\Delta\omega$, 
the phase mismatch is described by 
\begin{align}
\Delta k \sim \frac{\Delta\omega}{c} 
\left(
n(\lambda\u{c0}) - n(\lambda\u{p0}) 
- \lambda\u{c0}\ \od{n}{\lambda}\Big|_{\lambda=\lambda\u{c0}}
+ \lambda\u{p0}\ \od{n}{\lambda}\Big|_{\lambda=\lambda\u{p0}}
\right) 
\label{eq:mismatch}
\end{align}
in the first-order approximation with respect to $\Delta\omega$,
with rewriting the refractive-index terms into a wavelength-dependent form to clarify the correspondence with the Sellmeier equation. 
When $\lambda\u{p0}=\lambda\u{c0}=2\lambda\u{s0}$ is satisfied, 
the QPM condition $\Delta k = 0$ is still preserved even if the pump frequency is shifted by $\Delta\omega$. 
Therefore, in the region where linear approximation with respect to $\Delta\omega$ is valid, 
the combination of $\lambda\u{c0}$ and $\lambda\u{p0}$ with \SI{1560}{nm} as the midpoint keeps $\Delta k$ small,
leading to the broad phase matching regardless of material-dependent refractive indices. 

In contrast to the behavior of the first peak, 
the other peak observed at $\lambda\u{s0}=$\SI{934}{nm},
with the condition of $\lambda\u{c0}=\SI{1540}{nm}$ and $\lambda\u{p0}\sim\SI{2350}{nm}$, 
arises from a delicate balance of the dispersion terms of the device.
The refractive index used in our simulation decreases monotonically with respect to $\lambda$~\cite{Jundt1997}.  
As a result, in Eq.~(\ref{eq:mismatch}), 
$n(\lambda\u{c}) - n(\lambda\u{p}) > 0$ and $-\lambda\u{c}\U{d}n/\U{d}\lambda|_{\lambda=\lambda\u{c}} > 0$ are satisfied
for $\lambda\u{c} < \lambda\u{p}$. 
On the other hand, because the last term is negative, $\Delta k\sim 0$ can be satisfied.
In the present case, such a situation happens to be realized 
around $\lambda\u{s0}=$\SI{934}{nm} for a fixed wavelength of $\lambda\u{c0}=\SI{1540}{nm}$. 
We note that the results are highly robust against the choice of refractive index models. 
For example, nearly identical results are obtained for the Sellmeier equations reported in Refs.~\cite{Zelmon1997,Deng2006}. 

\section{Experimental setup}
\begin{figure}[t]
 \begin{center}
   \scalebox{0.11}{\includegraphics{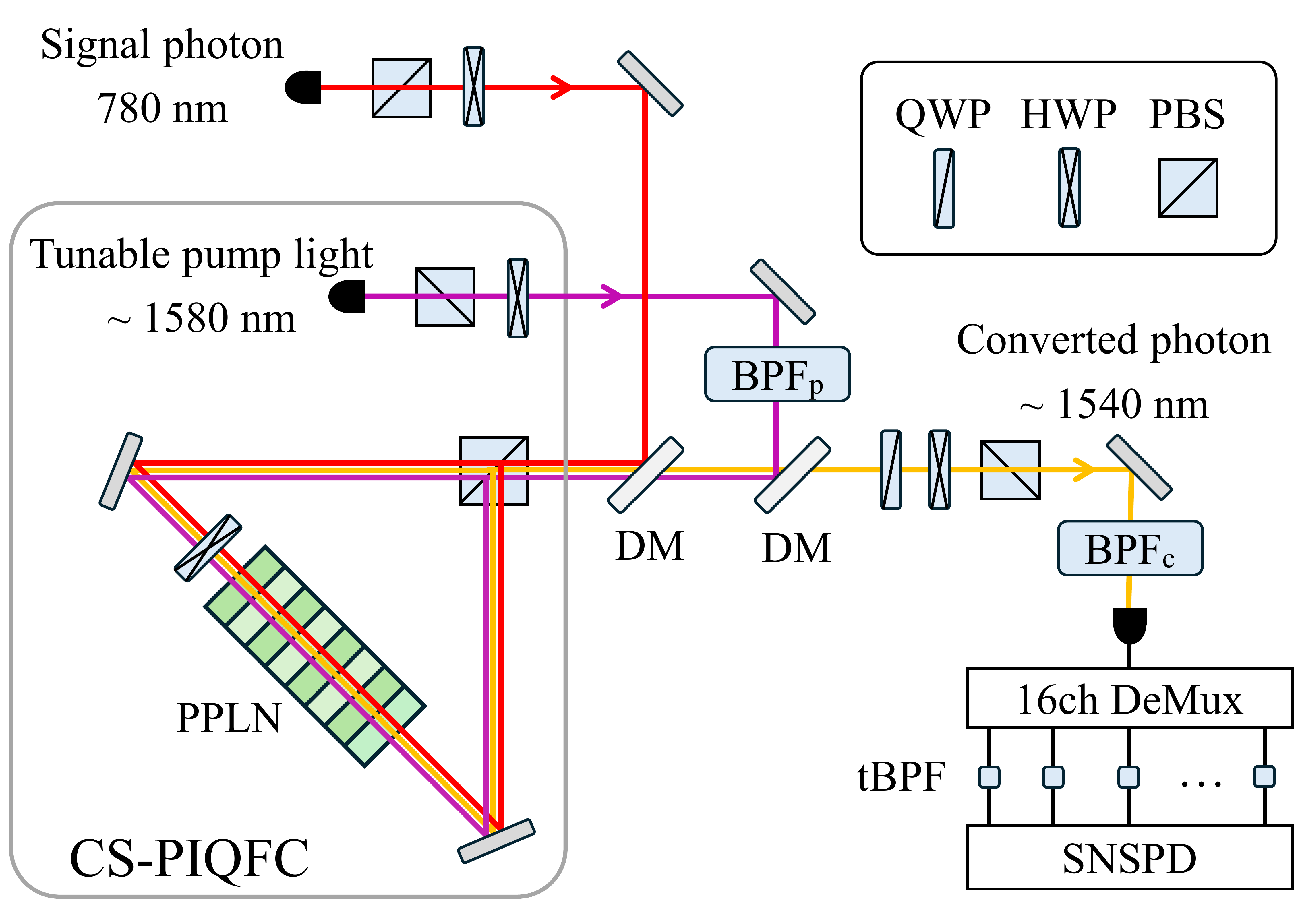}}   
      \caption{
        Experimental setup. The channel spacing of the DeMux is \SI{25}{GHz}. 
    }
    \label{fig:setup}
 \end{center}
\end{figure}
The experimental setup for a CS-PIQFC for quantum network hubs is shown in Fig.~\ref{fig:setup}.
The \SI{780}{nm} signal photon to be frequency-converted is prepared by spontaneous parametric downconversion~(SPDC) in a PPLN waveguide~(not shown).
A photon pair at \SI{780}{nm} and \SI{1541}{nm} is generated in the PPLN waveguide pumped by \SI{517}{nm} light.
The signal photon at \SI{780}{nm} is heralded by detecting the \SI{1541}{nm} photon,
coupled into a polarization-maintaining fiber~(PMF), and then coupled out to free space.
As in Fig.~\ref{fig:setup}, after being prepared in an appropriate polarization state 
using a polarizing beamsplitter~(PBS) followed by a half-wave plate~(HWP), and a quarter-wave plate~(QWP), 
the signal photon is coupled into the CS-PIQFC. 
The pump light used in the CS-PIQFC is prepared by a wavelength tunable laser, 
with the tuning range of 1572.063 to \SI{1607.760}{nm}. 
After amplification with an erbium-doped fiber amplifier~(EDFA, not shown) and coupled out into free space, 
the intensity ratio between the horizontal~(H) and vertical~(V) polarization components is adjusted using an HWP placed after a PBS. 
The pump light is spectrally filtered with a \SI{10}{nm} bandpass filter~($\U{BPF}\u{p}$) to suppress the broadband amplified spontaneous emission~(ASE) noise from the EDFA, 
and is then combined with the \SI{780}{nm} photon by a dichroic mirror~(DM) 
before being sent into the PIQFC. 

At the CS-PIQFC, the PBS separates the H- and V-polarized components into the counterclockwise~(CCW) and clockwise~(CW) directions, respectively. 
Using an HWP that flips H and V polarizations inside the interferometer,
all the light is coupled into the PPLN waveguide with V polarization for both directions. 
In the \SI{40}{mm}-long PPLN waveguide, 
the DFG process generates V-polarized photon around \SI{1540}{nm}. 
For the CCW and CW directions, V- and H-polarized converted photons are extracted from the interferometer, respectively,
with their polarizations inverted relative to the input.

When the quantum state of the signal photon at \SI{780}{nm} is $\alpha\ket{H}+\beta\ket{V}$, 
the extracted photon at \SI{1540}{nm} ideally becomes 
\begin{align}
\beta\sqrt{\eta\u{cw}}\ket{H}
+ \alpha\sqrt{\eta\u{ccw}} e^{i(\Delta\theta + \phi\u{p})}\ket{V},
\label{eq:state}
\end{align}
where $\eta\u{cw}$ and $\eta\u{ccw}$ are frequency conversion efficiencies for the CW and CCW directions, respectively, and 
$\Delta\theta = \theta\u{s}-\theta\u{p}-\theta\u{c}$ is the phase difference among the signal, pump and converted light introduced by the optical circuit. $\phi\u{p}$ is the relative phase between the H- and V-polarized pump light.
Under the conditions $\eta\u{cw}=\eta\u{ccw}$ and $\Delta\theta+\phi\u{p}=0$,
the QFC ideally implements a bit-flip~(Pauli $X$) operation. 
The bit flip can be deterministically corrected by applying an additional $X$ operation after the QFC,
thereby fully preserving the input quantum information. 

The quantum state of the converted photon is analyzed using a QWP, an HWP and a PBS. 
After a $\U{BPF}\u{c}$ with a bandwidth of \SI{10}{nm} for roughly suppressing the pump light, 
the converted photon is coupled to a single-mode fiber and distributed into 16 frequency channels 
using a DeMux with 25-GHz channel spacing aligned to the ITU-T DWDM grid,
labeled port 1~(\SI{194.850}{THz}) to port 16~(\SI{194.475}{THz}),
corresponding to \SI{1538.66}{nm} to \SI{1541.63}{nm}. 
The converted photon for each frequency channel is further frequency filtered by a tunable bandpass filter~(tBPF) 
with a bandwidth of \SI{0.075}{nm} and detected by a superconducting nanostrip single-photon detectors~(SNSPD). 
Each SNSPD has a detection efficiency of $\sim$ \SI{80}{\%}, 
a timing jitter of $\sim$ \SI{100}{ps}, and a dark count rate of $\sim$\SI{100}{cps}. 

\section{Experimental results}
\begin{figure}[t]
  \centering
  \scalebox{1}{\includegraphics{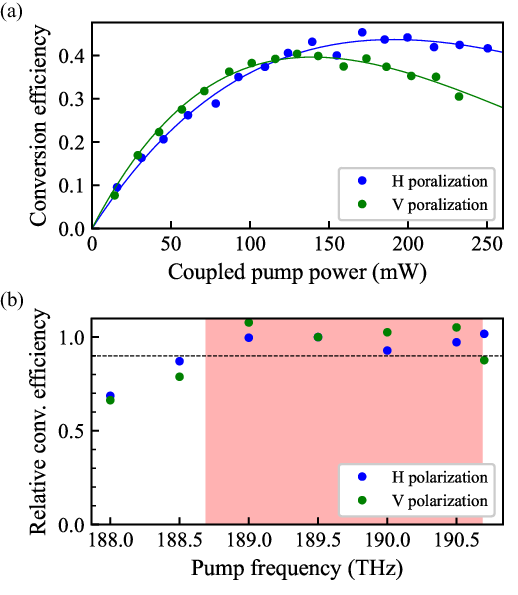}}    
    \caption{
      (a) Conversion efficiencies $\eta\u{ccw}$ and $\eta\u{cw}$ for H- and V-polarizaion
      including the coupling efficiencies of the \SI{780}{nm} light to the PPLN waveguide. 
      (b) Dependency of the maximum conversion efficiencies on the pump frequencies.
    }
  \label{fig:conv}
\end{figure}
Before demonstrating the CS-PIQFC using single photons, 
we measured the conversion efficiencies $\eta\u{ccw}$ and $\eta\u{cw}$ 
for H- and V-polarization, using a laser light at \SI{780}{nm}. 
An example of the conversion efficiencies when the pump frequency is \SI{189.5}{THz}~(\SI{1582.02}{nm})
corresponding to port 7 is shown in Fig.~\ref{fig:conv}~(a). 
From the the best fit of the experimental data 
to a function of $\eta\u{max}\sin^2(\sqrt{\eta\u{nor}P})$ gives 
$(\eta\u{max},\eta\u{nor})=(\SI{0.44},\SI{0.013}{mW^{-1}})$ and $(\SI{0.40},\SI{0.018}{mW^{-1}})$ for CCW and CW directions, respectively. 
The difference between the values for the CCW and CW directions can be compensated by properly tuning the intensity ratio of the H- and V-polarized components of the pump light to equalize the conversion efficiencies. 
The relative maximum conversion efficiencies for various pump frequencies are shown in Fig.~\ref{fig:conv}~(b). 
While the pump power for achieving the maximum conversion efficiency is not exactly equal for the directions and pump wavelengths,
we use a pump power of $\sim$ \SI{125}{mW} for each direction in this experiment. 
Over a range of about \SI{2}{THz}, from 188.7 to \SI{190.7}{THz}, 
the relative conversion efficiencies are over 0.9, 
which is comparable to the result reported in Ref.~\cite{Arizono2026}. 

\begin{figure}[t]
    \scalebox{0.55}{\includegraphics{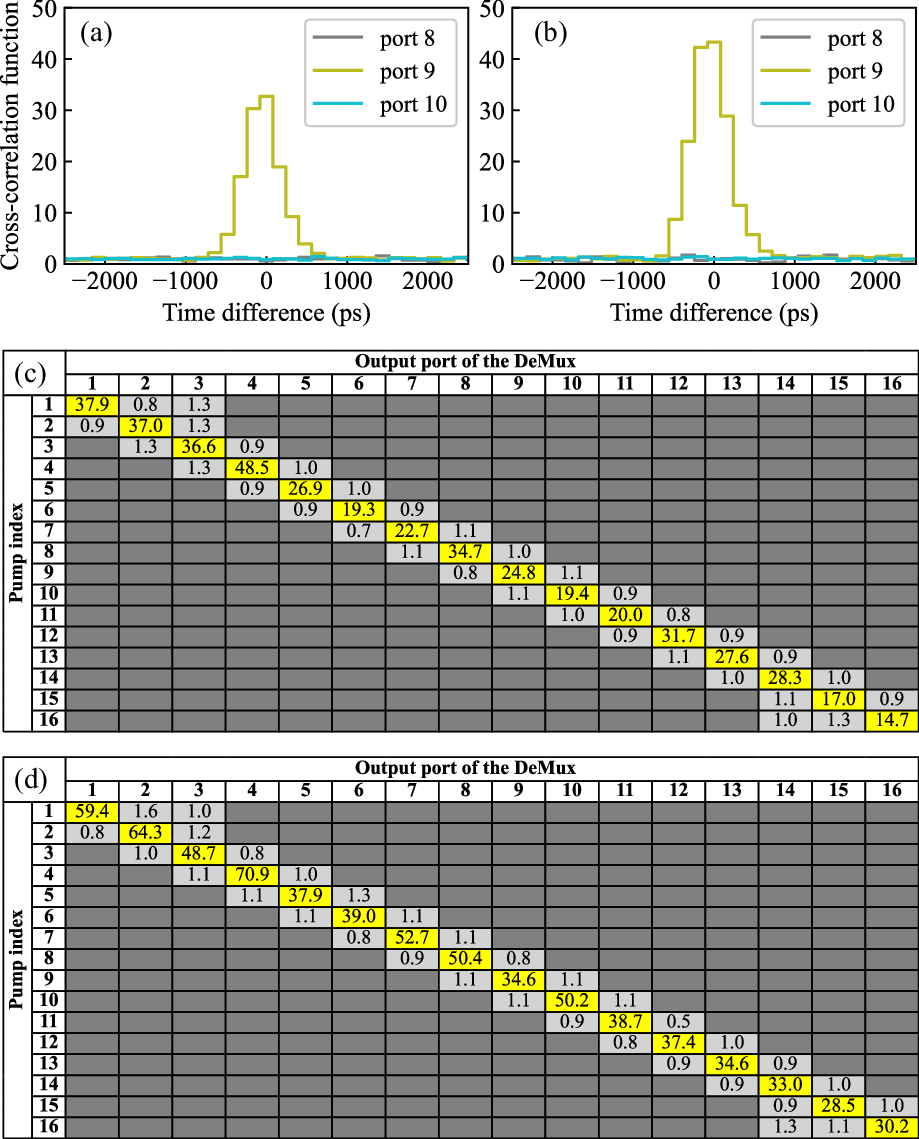}}
    \caption{Cross-correlation functions after CS-QFC 
      of output ports 8, 9 and 10 of DeMux for (a) CCW~(H) and (b) CW~(V) directions,
      using the pump frequency corresponding to port 9. 
      (c) and (d) are list of the cross correlation functions for CCW and CW directions, respectively. 
    }
  \label{fig:g2}
\end{figure}
Next, we demonstrate the CS-PIQFC using single photons. 
Without the QFC, the cross-correlation function between the initial paired photons 
at \SI{780}{nm} and \SI{1541}{nm} generated by SPDC is approximately 166. 
For the 16 different pump frequencies corresponding to the 16-channel DeMux installed in the path of the converted photons, we measured the cross-correlation functions after CS-PIQFC for all channels. 
Examples of the cross-correlation functions for CCW and CW directions
related to output ports 8, 9 and 10 of DeMux, obtained using the pump light corresponding to port 8, 
are shown in Fig.~\ref{fig:g2}~(a). 
The results show a highly nonclassical correlation between the photons after QFC in the desired channel.
In contrast, only nearly uncorrelated noise is observed in the neighboring channels, confirming the absence of crosstalk.
We also measured the cross-correlation functions for 16 main channel pairs and their neighboring channels.
For all 16 channels, nonclassical cross-correlation values greater than 14 are obtained 
with negligible crosstalk to the other channels, as listed in Figs.~\ref{fig:g2}~(c) and (d),  
indicating that the frequency-channel selection works as expected. 

\begin{figure}[t]
  \centering
  \scalebox{1}{\includegraphics{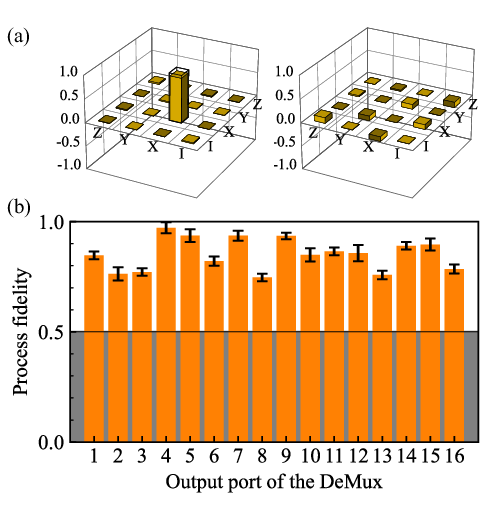}}
  \caption{
      (a) Real~(left) and imaginary~(right) parts of the reconstructed process matrix of CS-QFC to output port 9 of DeMux. 
      (b) Process fidelities for all output channels.  
    }
  \label{fig:process}
\end{figure}
The results presented so far were obtained by measuring the H- and V-polarized components separately. 
The PIQFC is designed to preserve an arbitrary polarization state of a photon after QFC. 
To verify this property, by preparing the input photon in polarization states $\ket{H}$, $\ket{V}$, $\ket{D}(=(\ket{H}+\ket{V})/\sqrt{2})$, and $\ket{R}(=(\ket{H}+i\ket{V})/\sqrt{2})$, we performed quantum state tomography for each input state. 
From the experimental results, we reconstructed process matrix $\chi\u{ex}$ of PIQFC. 
The reconstruction was performed for 16 QFC processes with different target conversion wavelengths, corresponding to port 1 to port 16, where the phase $\Delta \theta +\phi\u{p}$ in Eq.~(\ref{eq:state})
was calibrated to zero prior to the measurement of port 1 and was not adjusted thereafter. 
The ideal process matrix has only the diagonal element corresponding to $X=\ketbra{H}{V}+\ketbra{V}{H}$ equal to unity,
while all other elements are zero.
As an example, $\chi\u{ex}$ for port 9 is shown in Fig.~\ref{fig:process}~(a).
As expected, the process matrix exhibits a dominant $X$ component. 
The process fidelity with respect to this ideal process determined by $\tr(\chi\u{ex}X)$ is $0.94\pm 0.02$. 
In the same manner, we evaluated the process fidelities for the other 15 frequency channels
using the reconstructed process matrices. 
The results are summarized in Fig.~\ref{fig:process}. 
The minimum value among all channels is 0.75, 
which clearly exceeds the classical limit of 0.5~\cite{Massar1995,Nielsen2002}. 
Since the phase $\Delta\theta +\phi\u{p}$ and the pump power setting were not recalibrated for measurements across all frequency channels, the results confirm that our PIQFC operates reliably and with high fidelity regardless of the selected frequency channel. 

\section{Discussion}

We have experimentally demonstrated that
the CS-PIQFC from \SI{780}{nm} to the telecom wavelengths maintains high fidelity and efficiency 
across 16 frequency channels with the channel spacing of \SI{25}{GHz}. 
The performance of QFC is expected to be preserved over a wide tunable range of \SI{2}{THz}
in terms of the signal-to-noise ratio~\cite{Arizono2026}. 
Here, we discuss possible approaches and the feasibility of interfacing diverse quantum systems with the 780-nm hub.
For short-wavelength photons far from \SI{780}{nm},
suitable pump preparation for QFCs would be relatively easy, enabling connection to the hub. 
As examples, QFCs of \SI{422}{nm} photon for Sr$^+$ and \SI{493}{nm} photon for Ba$^+$ to \SI{780}{nm} require 
pump light at wavelengths of $\sim$ \SI{920}{nm} and $\sim$ \SI{1340}{nm}~\cite{Hannegan2021,Saha2023}, respectively. 
On the other hand, there exist quantum systems, such as NV center~(\SI{637}{nm}) and Ca$^+$~(\SI{854}{nm}), 
in wavelength regions that are not well suited to the quantum hub due to the small tunability,
and at the same time, preparing appropriate pump light for QFC to \SI{780}{nm}. 
For such quantum systems, 
connection to the hub can be facilitated by employing a two-stage QFC, first to the telecom band and then to \SI{780}{nm}. 
Alternatively, one may assign a dedicated telecom channel to such a quantum system via QFC, 
and use the hub to convert another photons between other quantum systems and this telecom channel like Refs.~\cite{Maring2017,Craddock2019}
for interactions such as Bell-state measurements. 

\begin{figure}[t]
 \begin{center}
    \scalebox{1}{\includegraphics{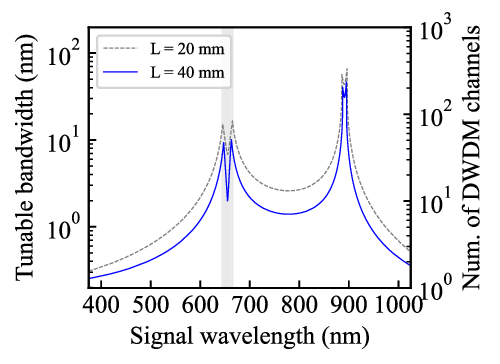}}
    \caption{
    Tunable bandwidth vs wavelength of the input photon 
    when the center of the target wavelength after QFC is \SI{1310}{nm}. 
    The bandwidth is evaluated assuming a \SI{20}{nm} separation between the pump and converted wavelengths.
    Channel spacing of the DWDM is \SI{25}{GHz}. 
    }
    \label{fig:Oband}
 \end{center}
\end{figure}
So far, we have focused our analysis on the telecom C band around \SI{1540}{nm}.
However, wavelengths in the telecom O band around \SI{1310}{nm} are also promising for quantum communication 
due to the expectation of high-fidelity integration with classical signals,
and related experimental works have been reported~\cite{Bock2018,Knaut2024,Liu2024,Thomas2024,Lu2026}. 
By performing the same analysis as in Fig.~\ref{fig:Cband}, 
when the center of the target telecom wavelength of CS-QFC is \SI{1310}{nm}, 
the tunable bandwidth is calculated as shown in Fig.~\ref{fig:Oband}. 
A trend similar to Fig.~\ref{fig:Cband} is observed, although the peak wavelengths are different. 
A peak around \SI{655}{nm} corresponds to the second harmonic of \SI{1310}{nm},
and thus the wideband property is material-independent. 
The pump light for CS-QFC is in the telecom O-band, which is advantageous for implementation. 
On the other hand, the other peak around \SI{900}{nm} is relatively sensitive to the refractive index 
and requires pump light around \SI{2875}{nm}. 

\section{Conclusion}
In conclusion, we have proposed a QFC-based network hub for bridging local quantum devices 
that emit photons at unique wavelengths and DWDM fiber networks. 
Due to the dispersion sweet spot arising from the \SI{780}{nm} band being located near the second harmonic of telecom C-band wavelengths,
we showed that this wavelength configuration inherently provides a broad phase-matching bandwidth.
Following the same principle, a hub wavelength around \SI{655}{nm} is suitable 
in terms of tunable bandwidth for a QFC-based network hub connecting to the the \SI{1310}{nm} telecom O-band. 
In the experiment of CS-PIQFC from \SI{780}{nm} to \SI{1540}{nm}, we achieved a wide tunable bandwidth of \SI{2}{THz}. 
We confirmed that by tuning the pump wavelengths,
the CS-PIQFC successfully distributed single photons to 16 frequency channels around \SI{1540}{nm}
via a DeMux with a \SI{25}{GHz} channel spacing on the ITU-T DWDM grid, yielding process fidelities of over \SI{75}{\%} in every channel.

In synergy with various other reconfigurable network functionalities and/or hub-based network architectures without QFC~\cite{Monroe2014,Dong2023,Sakuma2024,Chen2026,Huang2026,Pouryousef2026}, the QFC-based network hub will be a flexible backbone capable of interconnecting diverse quantum devices, including both photonic and matter systems, over frequency-multiplexed telecom networks.
While the present study identifies the \SI{780}{nm} band as an optimal hub wavelength based on conventional PPLN waveguides, this concept can be broadly extended. 
Exploring alternative platforms, such as thin-film lithium niobate~\cite{Han2021,Wang2023} and periodically poled potassium titanyl phosphate~(PPKTP)~\cite{Mann2023}, as well as chirped poling techniques~\cite{Suchowski2008,Chen2025,Fang2026},
holds the potential for designing QFC-based network hubs operating at other wavelengths. 
We believe that the insights presented here provide a valuable foundation for future work. 

\section*{Acknowledgements}
We thank Shoichi Murakami for assistance with the experiment.
We also thank Toshiki Kobayashi for fruitful discussions. 
T.Y. and R.I. acknowledge the members of the Quantum Internet Task Force for the comprehensive and interdisciplinary discussions on the quantum internet. 
This work was supported by JST FOREST Program~(JPMJFR222V); 
R \& D of ICT Priority Technology~(JPMI00316);
JST Moonshot R \& D~(JPMJMS2066, JPMJMS226C); 
MEXT/JSPS KAKENHI~(JP25K01263).

\end{document}